\documentstyle[mprocl,psfig,graphicx]{article}

\def\be{\begin{equation}}
\def\ee{\end{equation}}
\def\bea{\begin{eqnarray}}
\def\eea{\end{eqnarray}}

\begin{document}

\title{FACING NON-STATIONARY CONDITIONS WITH A NEW INDICATOR OF ENTROPY INCREASE: THE CASSANDRA ALGORITHM}




\author{P. ALLEGRINI}

\address{ Istituto di Linguistica Computazionale del Consiglio Nazionale delle
Ricerche,\\
 Area della Ricerca di Pisa--S. Cataldo, Via Moruzzi 1, 
56124, Pisa, Italy}

\author{P. GRIGOLINI}
\address{Center for Nonlinear Science, University of North Texas,
P.O. Box 5368,\\ Denton, TX 76203;\\
Istituto di Biofisica del Consiglio Nazionale delle
Ricerche,\\
 Area della Ricerca di Pisa--S. Cataldo, Via Moruzzi 1, 
56124, Pisa, Italy}

\author{P. HAMILTON}
\address{Center for Nonlinear Science, Texas Woman's University, 
P.O. Box 425498, Denton, Texas 76204}

\author{L. PALATELLA, G. RAFFAELLI, M. VIRGILIO}

\address{Dipartimento di Fisica dell'Universit\`{a} di Pisa and
INFM 
Piazza
Torricelli 2, 56127 Pisa, Italy }


\maketitle\abstracts{We address the problem of detecting
non-stationary effects in time series (in particular fractal time
series) by means of the Diffusion Entropy Method (DEM). This means
that the experimental sequence under study, of size $N$, is explored
with a window of size $L << N$.  The DEM makes a wise use of the
statistical information available and, consequently, in spite of the
modest size of the window used, does succeed in revealing local
statistical properties, and it shows how they change upon moving the
windows along the experimental sequence.  The method is expected to
work also to predict catastrophic events before their occurrence.}

\section{Introduction}
The main aim of this paper is to illustrate a promising strategy to
study non-stationary processes. We prove that the method is efficient
by means of the joint study of real and artificial sequences, and
we reach a conclusion that makes it plausible to
imagine the method at work to successfully  predict
the time of occurrence of catastrophic events.

The method here illustrated is a suitable extension of the 
Diffusion Entropy Method (DEM). The DEM is discussed in details in 
other publications\cite{nic1,nic2,giacomo}. 
Here we limit ourselves to give a concise 
illustration of this technique so as to allow the reader to 
understand the spirit of the method of this paper, at least through a 
first reading, without consulting these earlier publications. 
 The first step of this
technique is the same as that of the pioneering work of
Refs. \cite{detrendingindna,detrendinginheartbeating}. 
This means that
the experimental sequence is converted into a kind of Brownian-like
trajectory. The second step aims at deriving many distinct diffusion
trajectories with the technique of moving windows of size $l$. The
reader should not confuse the mobile vindow of size $l$ with the
mobile window of value $L$ that will be used later on in this paper to
detect non-stationary properties. For this reason we shall refer to
the mobile windows of size $L$ as large windows, even if the size of
$L$ is relatively small, whereas the mobile windows of size $l$ will
be called small windows. The large mobile window has to be
interpreted as a sequence with statistical properties to reveal, and
will be analyzed by means of small windows of size $l$, with $l <
L$. The success of the method depends on the fact that the DE makes a
wise use of the statistical information available. In fact, the small
windows overlap among themselves and are obtained by locating their left border on the
first site of the sequence, on the second second site, and so on. 
The adoption of overlapping windows is dictated by the wish to 
establish a connection with the Kolomogorov-Sinai (KS)\cite{beck,dorfman}
method. This yields also, as a further beneficial effect, many more 
trajectories than the widely used method of Detrended Fluctuation 
Analysis (DFA)\cite{detrendingindna,detrendinginheartbeating}.

In conclusion, we create a conveniently large number of trajectories by
gradually moving the small window from the first position, with the left border
of the small window coinciding with the first size of the sequence,
to the last position, with the right border of the small window
coinciding with the last site of the sequence. Atfter this stage, we
utilize the resulting trajectories, all of them with the initial position
located at $x = 0$, to produce a probability distribution at
``time'' $l$. We evaluate the Shannon entropy of this distribution,
$S_{d}(l)$,
and with easy mathematical arguments we prove that, if the diffusion
process undergoes a scaling of intensity $\delta$, then
\begin{equation}
S_{d}(l) = A + \delta ln (l).
\label{shapeofdiffusionentropy}
\end{equation}
Thus the parameter $\delta$ of the scaling condition, if this condition 
applies, can be measured without
recourse to any form of detrending.  

This is the original motivation for the DEM\cite{nic1,nic2,giacomo}. 
In this paper we 
want to prove that the DEM does much more than detecting scaling. In 
a stationary condition the DEM not only detects with accuracy the 
final scaling but it also affords a way to monitor the regime of 
transition to the final \emph{thermodynamic} condition. If the 
sequence under study is affected by biases and non-stationary 
perturbations, the attainment of the final regime of steady scaling 
can be cancelled, and replaced by an \emph{out of equilibrium} regime 
changing in time under the influence of time dependent biases. We 
want to prove that the DEM can be suitably extended to face this 
challenging non-stationary condition.

The outline of the paper is as follows. In Section 2 we 
shortly review the main tenets of the DEM so as to make this paper, 
as earlier mentioned, as self-contained as possible. In Section 3 
we illustrate the extension of the DEM  and we discuss the 
fundamental problem of assessing which is the shortest portion, of 
size $L$, of the sequence under study, which is still large enough as 
to  make the DEM work. In Section 3 we express our hope that this 
new method
might serve prediction purposes.

\section{Diffusion Entropy}\label{II}

Let us consider a sequence of $M$ numbers $ \xi_{i}$ , with  $i = 
1,  \ldots , M$.
The purpose of the DEM algorithm is to establish the possible
existence of a scaling, either normal or anomalous, in the most
efficient way as possible, without altering the data with any form
of detrending. Let us select first of all an integer number
$l$, fitting the condition $1 \leq l \leq M$.
As earlier mentioned, we shall refer ourselves to $l$ as "time".
For any given time $l$ we can find $M - l +1$ sub-sequences defined by
\begin{equation}
\xi_{i}^{(s)} \equiv \xi_{i + s}, \quad   \quad s = 0, 
\ldots ,  M-l.
\label{multiplicationofsequence}
\end{equation}
For any of these sub-sequences we build up a diffusion
trajectory, labelled with the index
$s$, defined by the position
\begin{equation}
x^{(s)}(l) = \sum_{i = 1}^{l} \xi_{i}^{(s)}
= \sum_{i = 1}^{l} \xi_{i+s}.
\label{positions}
\end{equation}

Let us imagine this position as referring to a Brownian 
particle that at regular intervals of time has been jumping forward or
backward according to the prescription of the corresponding
sub-sequence of Eq.(\ref{multiplicationofsequence}). This means
that the particle before reaching the position that it holds at
time $l$ has been making $l$ jumps. The jump made at the
$i$-th step has the intensity $|\xi_{i}^{(s)}|$ and is forward or
backward according to whether the number $\xi_{i}^{(s)}$ is
positive or negative.

We are now ready to evaluate the entropy of this diffusion process.
To do that we have to partition the $x$-axis into cells of size
$\epsilon(l)$. When this partition is made we have to label the
cells. We count how many particles are found in the same cell at a
given time $l$. We denote this number by $N_{i}(l)$. Then
we use this number to determine the probability that a particle
can be found in the $i$-th cell at time $l$, $p_{i}(l)$, by means of
\begin{equation}
p_{i}(l) \equiv  \frac{N_{i}(l)}{(M-l+1)} .
\label{probability}
\end{equation}
At this stage the entropy of the diffusion process at the time $l$
is determined and reads
\begin{equation}
S_{d}(l) = - \sum_{i} p_{i}(l) ln [p_{i}(l)].
\label{entropy}
\end{equation}
The easiest way to proceed with the
choice of the cell size, $\epsilon(l)$, is to assume it independent
of $l$ and determined by a suitable fraction of the square root of
the variance of the
fluctuation $\xi(i)$.

Before proceeding with the illustration of how the DEM method works,
it is worth making a comment on the way we use to define the trajectories.
The method we are adopting is based on the idea of a moving window of
size $l$ that
makes the $s-th$ trajectory closely correlated to the next, the
$(s+1)-th$ trajectory. The two trajectories have $l-1$ values in
common. 
A motivation for our choice 
is given by our wish to establish a
connection with the Kolmogorov Sinai (KS) entropy\cite{beck,dorfman}.
The KS entropy of a symbolic sequence is evaluated by moving a window
of size $l$ along the sequence. Any window position corresponds to a
given combination of symbols, and, from the frequency of each
combination, it is possible to derive the Shannon entropy $S(l)$. The
KS entropy is given by the asymptotic
limit $lim_{l \rightarrow \infty} S(l)/l$. We believe that the same
sequence, analyzed with the DEM method, at the large values of $l$ 
where $S(l)/l$ approaches the KS value,
 must yield a well defined scaling $\delta$. To realize this
correspondence we carry out the determination of the Diffusion Entropy using the
same criterion of overlapping windows as that behind the KS entropy.

Details on how to deal with the transition from the
short-time regime, sensitive to the discrete nature of the
process under study, to the long-time limit where both space
an time can be perceived as continuous, are given in
Ref. \cite{nic1,giacomo}. 
Here we make the simplifying assumption
of considering so large times as to make the continuous assumption
valid.
In this case the trajectories, built up with the above illustrated procedure,
correspond to the following equation of motion:

\begin{equation}
\frac{dx}{dt} = \xi(t) ,
\label{equationofmotion}
\end{equation}
where $\xi(t)$ denotes the value that the time series under study
gets at the $\l-th$  site. This means
that the function $\xi(l)$ is
depicted as a function of $t$, thought of as a continuous time $t = l$.
In this case the Shannon entropy reads
\begin{equation}
S_{d}(t) = - \int_{-\infty}^{\infty} dx \, p(x,t) ln [p(x,t)].
\label{continuousshannonentropy}
\end{equation}
We can derive with a simple treatment an analytical solution for Diffusion Entropy when the process is characterized by {\em scaling}, namely when
\begin{equation}
p(x,t) = \frac{1}{t^{\delta}} \, F\left( 
\frac{x}{t^{\delta}}\right).
\label{rigorousdefinition}
\end{equation}
Let us plug Eq.(\ref{rigorousdefinition}) into
Eq.(\ref{continuousshannonentropy}). After a simple algebra, we get:
\begin{equation}
S_{d}(\tau) = A + \delta(\tau) \tau ,
\label{keyrelation}
\end{equation}
where
\begin{equation}
A \equiv -\int_{-\infty}^{\infty} dy \, F(y) \, ln [F(y)]
\label{ainthecontinuouscase}
\end{equation}
and
\begin{equation}
\tau \equiv \ln (t) .
\label{logarithmictime}
\end{equation}



It is evident that this kind of technique to detect scaling does
not imply any form of detrending, and this is one of the reasons why
some attention should be devoted to it. It is also worth mentioning,
as we prove now, that it yields the correct scaling values even for 
the so-called {\em L\'evy walks}, where the time dependence
of the second moment with respect to time has an exponent which
is different from the scaling exponent of the L\'evy process.

We therefore
check the efficiency of this technique by the studying
the artificial sequence of Refs. \cite{marcoluigi,allegro}. This
sequence is built up in such a way as to realize long sequences
of either $+$'s or $-$'s. The probability of finding a sequence
of only $+$'s or only $-$'s of length $t$ is given by
\begin{equation}
\label{chosenanalyticalform}
\psi(t) = (\mu - 1) \frac{T^{\mu-1}}{(t+T)^{\mu}}.
\end{equation}
Here we focus our attention on the condition $\mu < 3$ and we
raise the reader's attention on the interval $2 \leq \mu \leq 3$.
In fact, this kind of sequence is the same as that adopted
in earlier work \cite{allegro} for a dynamic derivation of L\'{e}vy
diffusion, which shows up when the
condition $2 < \mu < 3$ applies.
It corresponds to a particle travelling with constant velocity
throughout the whole time interval corresponding to either only +'s
or only -'s, and changing direction with no rest, at the end of any
string with the same symbols.

\begin{figure}[h]
\begin{center}
\centerline{\psfig{figure=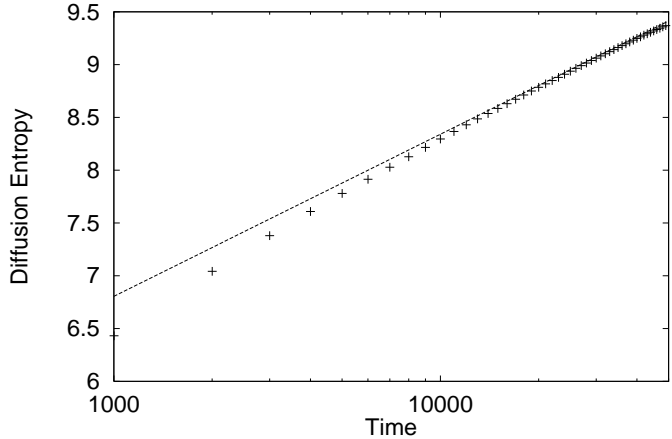,width=9cm}}
\caption{\label{SVMDE}
The diffusion entropy as a function of time. The numerical method is
applied to the artificial sequence described above, with $\mu = 2.5$,
studied according to the SVM prescription. According to the
theoretical arguments of the text the scaling parameter $\delta$ is the
slope of the straight line fitting the numerical results at large
times, which yields in this case $\delta = 2/3 = 1/(\mu -1)$ ($T = 10$).}
\end{center}
\end{figure}

We will refer to this model as {\em Symmetric Velocity Model} (SVM).
We know from the theory of Ref.\cite{allegro} that the scaling
of the resulting diffusion process when $2<\mu<3$ is
\begin{equation}
\delta = \frac{1}{\mu -1}.
\label{levyscaling}
\end{equation}
Note, however, that this diffusion process has a finite propagation
front, with ballistic peaks showing up at both $x = t$ and $x = -t$.
The intensity of these peaks is proportional
to the correlation function
\begin{equation}
\Phi_{\xi}(t) = \left ( \frac{T}{T+t} \right )^{\mu -2}.
\label{correlatonfunction}
\end{equation}
As a consequence of this fact, the whole distribution does not
have a single rescaling. In fact, the distribution enclosed
between the two peaks rescales with $\delta$ of
Eq.(\ref{levyscaling}) while the peaks are associated to $\delta =1$.
Furthermore, it is well known\cite{allegro} that the scaling of the second
moment is given by
\begin{equation}
\delta_{H} = \frac{4 - \mu}{2}.
\label{secondmomentscaling}
\end{equation}
Thus, it is expected that the scaling detected by the DE method
might not coincide with the prediction of Eq.(\ref{levyscaling})
for the whole period of time corresponding to the presence of peaks
of significant intensity. We think that the L\'{e}vy scaling of 
Eq.(\ref{levyscaling}) will show up at long times, when the peak
intensity is significantly reduced.
This conjecture seems to be supported
by the numerical results illustrated in Fig.\ref{SVMDE}. We see in fact that
the scaling predicted by Eq.(\ref{levyscaling}) is reached after an extended
transient, of the order of about 20,000 in the scale of Fig.\ref{SVMDE}.
This time interval is about 2000 larger than the value assigned to the
parameter T, of Eq.(\ref{chosenanalyticalform}),
which is, in fact, in the case of Fig.\ref{SVMDE}, T= 10.

In conclusion, this section proves that the DE method applied
to the SVM yields, for the scaling parameter $\delta$, the correct
value of Eq.(\ref{levyscaling}),
rather than the value that would be obtained
measuring the variance of the diffusion process, Eq.(\ref{secondmomentscaling}).
However, the time necessary to make this correct value emerge is very large.
Furthermore, as shown in Fig. 2, the adoption of
SVM  would make the scaling parameter $\delta$ insensitive to $\mu$ in
the whole interval $1 \leq \mu \leq 2$. This means that the adoption of the DE
method would not allow us
to distinguish a process with $\mu$ very close to 1 from one with $\mu$
very close to 2. This problem can be solved using different rules
for the diffusion process \cite{giacomo}: the random walker can,
for instance, walk always in the same direction, and at the ``time''
when there is a passage from a laminar region of \+'s to one
of -'s and vice-versa. If this latter rule is adopted, then
it is easy to prove \cite{giacomo} that the resulting $p(x,t)$
is an asymmetric L\'evy distribution, with a scaling $\delta = \mu-1$
for $1 \leq \mu \leq 2$.

Throughout this paper, however, apply the DEM with the rule corresponding
to the symmetric velocity model, with $\delta$ depending on $\mu$ as in
Fig. 2. In the regime of ordinary statistical 
mechanics ($\mu \gg 3$) the ordinary scaling is quickly attained, while
the condition of anomalous statistical mechanics $\mu < 3$ is 
characterized by a long transient regime, which is carefully recorded 
by the DEM. In this paper we want to use the DEM to monitor the time 
dependence of the ``rules" generating the sequences under study.

\begin{figure}[ht]
\label{regola2}
\centerline{\includegraphics[angle=270,scale=.50]{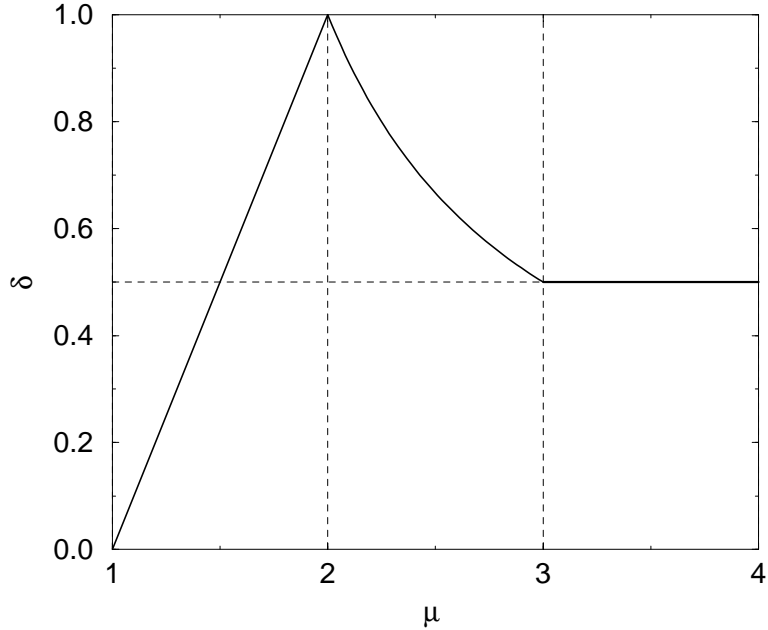}}
\caption{$\delta$ as a function of $\mu$ according to the prescriptions
of Ref. [3].} 
\end{figure}

\section{The new method at work with nonstationary sequences}
To illustrate the ideas that led us to propose the method of analysis 
with two moving window, let us begin with discussing the artificial 
sequence given by
\begin{equation}
\xi_{b}(t) = \kappa\xi(t) + A cos (\omega t).
\label{withbias}
\end{equation}
The second term on the right hand side of this equation is a
deterministic contribution that might mimic, for instance, the season
periodicity of Ref. [1].
The first term 
on the right 
hand side is a fluctuation with no correlation that can be correlated 
or not to the harmonic bias.

Fig. 3 refers to the case when the random fluctuation has no 
correlation with the harmonic bias. It is convenient to illustrate 
what happens when $\kappa = 0$. This is
the case 
where the signal is totally deterministic. It would be nice
if the entropy in this case did not increase upon increasing $l$.
However, we must notice that the method of mobile windows implies that
many trajectories are selected, the difference among them being, in
the determinist case where $\xi_{b}(t) = A cos (\omega t)$, a
difference on initial conditions. Entropy initially increases. This
is due to the fact that the statistical average on the initial conditions is
perceived as a source of uncertainty. However, after a time of the
order of the period of the deterministic process a regression to the
condition of vanishing entropy occurs, and it keeps repeatedly
occurring for the multiple times. Another very remarkable fact is
that the maximum entropy value is constant, thereby signalling
correctly that we are in the presence of a periodic signal, where the
initial entropy increase, due to the uncertainty on the initial
conditions, is balanced by the recurrences. 
Let us now consider the effect of a non vanishing $\kappa$. We see that 
the presence of an even very weak random component makes an abrupt 
transition to occur from the condition where the diffusion entropy is 
bounded from above, to a new condition where the recurrences are 
limited from below by an entropy increase proportional to $0.5 \ln l$. 
In the asymptotic time regime the DEM yields, as required, the proper 
scaling $\delta =0.5$. However, we notice that it might be of some 
interest for a method of statistical analysis to give information on 
the extended regime of transition to the final \emph{thermodynamic} 
condition. We notice that if the DEM is interpreted as a method of 
scaling detection, it might also give the impression that a scaling 
faster than the ballistic $\delta$ is possible. This would be 
misleading. However, this aspect of the DEM, if conveniently used, 
can become an efficient method to monitor the non-stationary nature 
of the sequence under study.

\begin{figure}
\centerline{\includegraphics[angle=270,scale=.50]{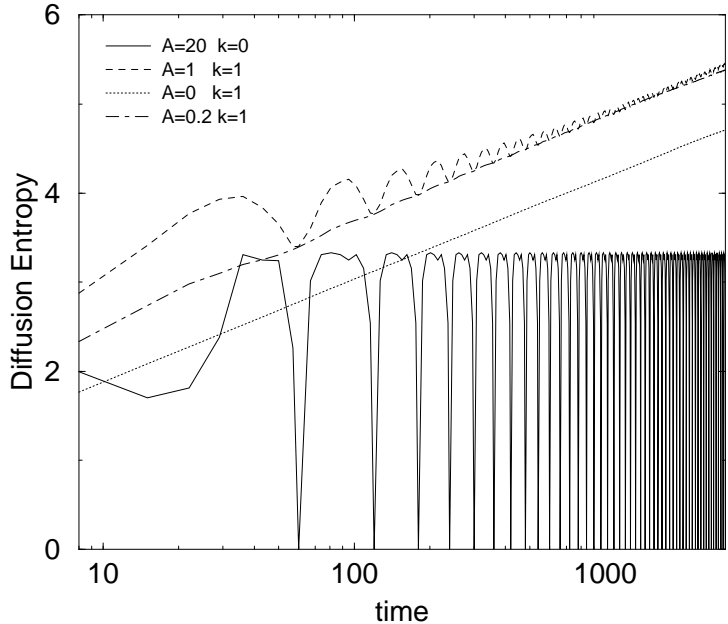}}
\caption{\label{fig1} The diffusion entropy $S_{d}(l)$ as a 
function of time $l$ for 
different sequences of the type of Eq. (12).} 
\end{figure}

\begin{figure}
\begin{center}
\centerline{\includegraphics[angle=270,scale=.40]{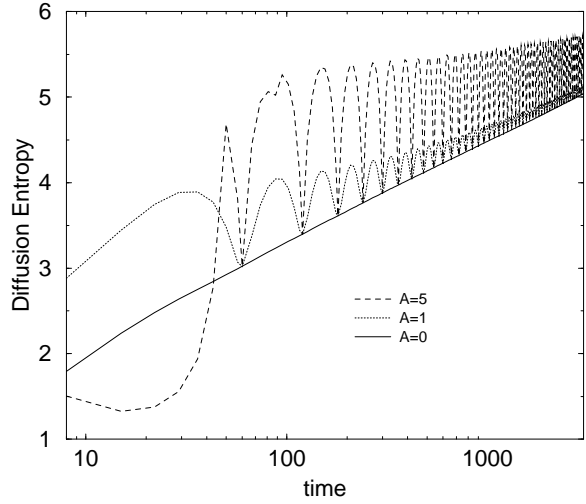}}
\caption{\label{fig2}
The diffusion entropy $S_{d}(l)$ as a function of time $l$ for 
different sequences of the type of Eq. (12) with the prescription of 
Eq. (13) for the random component.
} 
\end{center}
\end{figure}

In the special case where the fluctuation $\xi(t)$ is correlated or
anticorrelated to 
the bias, 
the numerical results illustrated in Fig. 4 show that 
the time evolution of the diffusion entropy is qualitatively similar 
to that of Fig. \ref{fig1}.
The correlation between the first and the second term on
the right hand side of Eq. (\ref{withbias}) is established by assuming
\begin{equation}
\xi(t) = \xi_{0}(t) cos (\omega t),
\label{modulatedintensity}
\end{equation}
where $\xi_{0}(t)$ is the genuine independent fluctuation, 
without memory, whose intensity is modulated to establish a
correlation with the second term.
It is of some interest to mention what happens when $A = 0,
\kappa = 1$, 
and consequently $\xi_{b}(t)$ coincides with
$\xi(t)$ of Eq. (\ref{modulatedintensity}). In this case we get
the straight (solid) line of Fig. \ref{fig2}. 
This means that the adoption of the
assumption that the process is stationary yields a result that is
independent of the modulation.

\begin{figure}
\label{sinus}
\begin{center}
\centerline{\includegraphics[angle=270,scale=.40]{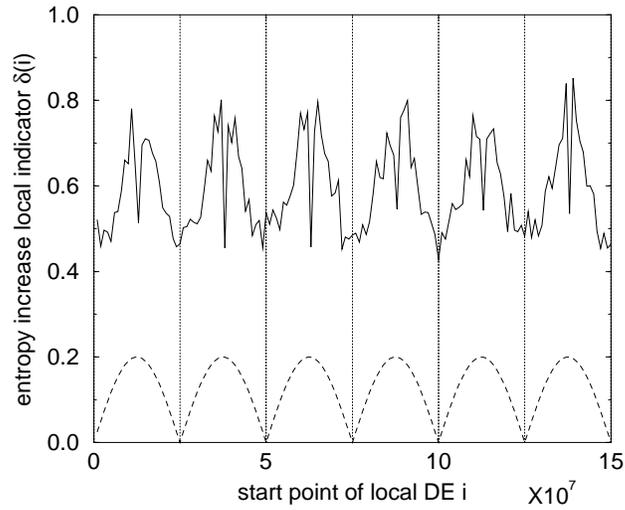}}
\caption{\label{fig3}
The method of the two mobile windows applied to a sequence 
given by Eq. (12) with $A = 0$ and $\xi(t)$ given by Eq. (13).
The dashed line represents the sinus' amplitude (not in scale) corresponding to the position $i$ of the left border of the large moving window. 
}
\end{center}
\end{figure}

We use this interesting case to illustrate the extension of the 
DEM, which is the main purpose of this paper. As earlier mentioned, 
this extension is based on the use of two mobiles windows, one of 
length L and the traditional ones of length $l \ll L$.
This
means that a large window of size $L$, with $L \ll T=2\pi/\omega$, is
located in a given position $i$ of the sequence under study, with $i
\leq N-L$, and the portion of the sequence contained within the window
is thought of as being the sequence under study. We record the
resulting $\delta$ (obtained with a linear regression method) and then
we plot it as a function of the position $i$. We show in Fig. \ref{fig3} that
this way of proceeding has the nice effect of making the periodic
modulation emerge.

Let us now improve the method to face non-stationary condition 
even further. As we have seen, the presence of time dependent 
condition tends to postpone or to cancel the attainment of a scaling 
condition. Therefore, let us renounce to using Eq. (9) and let us 
proceed as follows. For any large mobile window of size $L$ let us call 
$l_{max}$ the maximum size of the small windows. Let us call $n$ the 
position of the left border of the large window, and and let us 
evaluate the following property
\begin{equation}
I(n) \equiv \sum_{l=2}^{l_{max}} \frac{S_{d}(l) - [S_{d}(1)+0.5 \ln l]}{l}.
\end{equation}
The quantity $I(n)$ detects the deviation from the condition of 
increase that the diffusion entropy would have in the random case. 
Since in the regime of transition  the entropy increase can be much 
slower than in the corresponding random case, the quantity $I(n)$ can 
also bear negative values. This indicator affords a satisfactory way 
to detect local properties. As an example,
Fig. 6 shows a case based on the DNA model of
Ref. [8], 
called Copying Mistake Map (CMM). This is a sequence of symbols $0$ and $1$
obtained from
the joint action of two independent sequences, one equivalent
to tossing a coin and the other equivalent to establishing
randomly
a sequence of patches whose length is distributed as an inverse
power law with index $\mu$ fitting the condition $2 < \mu < 3$.
The probability of using the former sequence is $1 - \epsilon$
and that of using the latter is $\epsilon$.
We choose a time dependent value of $\epsilon$ 

\begin{equation}
\epsilon=\epsilon_0 [1-cos(\omega t)].
\end{equation}
 In Fig. 6 we show how this periodicity is
perceived by using the two-windows generalization, proposed 
in this paper, of the DEM.

\begin{figure}
\centerline{\includegraphics[angle=270,scale=.50]{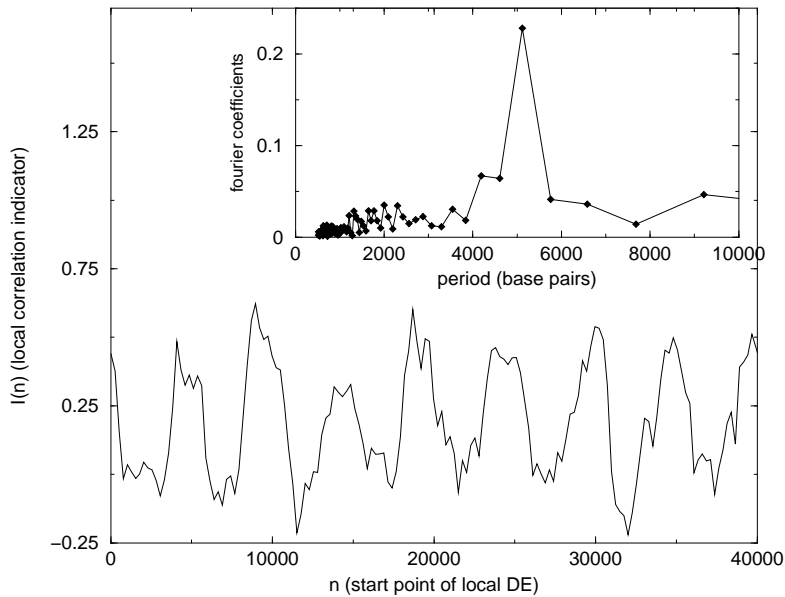}}
\caption{\label{fig5} The method of the two moving windows with $l_{max}=30$ applied
to the analysis of an artificial CMM sequence with periodic parameter
$\epsilon$. The period of the variation of $\epsilon$ is $5000$ bps and 
the analysis is carried out with moving windows of size $2000$ bps.
Inset: Fourier spectral analysis of $I(n)$.}

\end{figure}


As a final example to show the efficiency of the new method of 
analysis, let us address the problem of the search of hidden 
periodicities in DNA sequences.
Fig. 7 shows a
distinct periodic behavior for the human T-cell receptor alpha/delta
locus. A period of about 990 base pairs is very neat in
the first part of the sequence (promoter region), while
several periodicities of the order of $1000$ base pairs are distributed
along the whole sequence. 

These periodicities, probably due to DNA-proteins interactions in
active eukaryotic genes, are expected by biologists, but the current
technology is not yet adequate to deal with this issue, neither from
the experimental nor from the computational point of view: such a
behavior cannot be analyzed by means of crystallographic or structural
NMR methods, nor would the current (or of the near future) computing
facilities allow molecular dynamics studies of systems of the order of $10^6$
atoms or more.

\begin{figure}
\centerline{\includegraphics[angle=270,scale=.50]{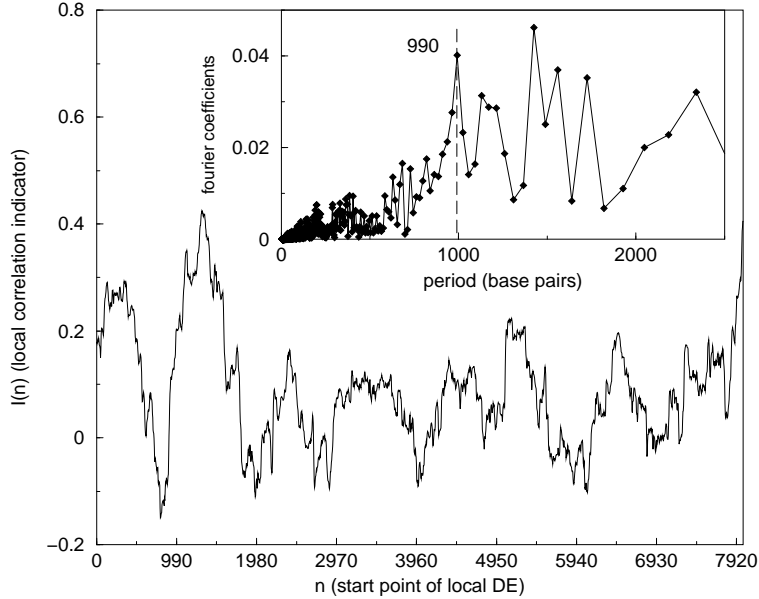}}
\caption{\label{fig4} 
The method of two mobile windows applied to the analysis of the human 
DNA sequence. The method of two mobile windows ($l_{max}=20$
$L=512$) detects a periodicity 
of 990 bps. Inset: Fourier spectral analysis of $I(n)$. 
} 
\end{figure}

\section{Conclusions}
The research work illustrated in this paper shows that the DEM is a 
very efficient way to detect the departure from ordinary Brownian 
motion with
the shortest sequence as possible. On the basis of these results we 
are confident that it will be possible to predict the occurrence of 
catastrophic events, heart-quakes, heart attacks, stock-market 
crashes, and so on. We think that if all these misfortune events are 
anticipated by a
correlation change, lasting for a fairly extended time period, then 
the DEM, within the double window procedure here illustrated, will 
signal in time their later occurrence. We refer to this method of 
analysis as Complex Analysis of Sequences via Scaling AND Randomness 
Assessment (CASSANDRA), and we hope to prove by means of future 
research work that its \emph{prophetic} power is worth of 
consideration. We wish that the CASSANDRA algorithm will have more 
fortune and will receive more credit than the daughter of Priam and 
Hecuba.


\end{document}